\documentclass[aps,prb,twocolumn,showpacs,superscriptaddress]{revtex4-1}
\usepackage{latexsym}
\usepackage{ifpdf}
\usepackage{graphicx}
\usepackage{epsfig}
\usepackage{amsmath, amssymb, bbm, mathrsfs}
\usepackage{multirow}
\usepackage[varg]{txfonts}
\usepackage{ulem}
\usepackage{fancyhdr}
\usepackage{color}
\usepackage{bm}
\usepackage{eurosym}
\usepackage{subfigure}
\usepackage{comment}
\usepackage[colorlinks=true]{hyperref}
\usepackage{scrhack}
\usepackage{physics}
\usepackage{balance}
\usepackage{float}
\usepackage{flushend}

\newcommand{\Ee}{{\rm e}}

\newcommand{\Ii}{{\rm i}}

\begin{document}
	\title{Strong coupling-enabled broadband non-reciprocity}
	\author{Xufeng Zhang}
	\email{xufeng.zhang@anl.gov}	
	\affiliation{Center for Nanoscale Materials, Argonne National Laboratory, Argonne, IL 60439, USA}
	\author{Alexey Galda}
	\affiliation{James Franck Institute, University of Chicago, Chicago, IL 60637, USA}
	\affiliation{Materials Science Division, Argonne National Laboratory, Argonne, IL 60439, USA}
	\author{Xu Han}
	\affiliation{Center for Nanoscale Materials, Argonne National Laboratory, Argonne, IL 60439, USA}
	\author{Dafei Jin}
	\affiliation{Center for Nanoscale Materials, Argonne National Laboratory, Argonne, IL 60439, USA}
	\author{V.\,M.\,Vinokur}
	\affiliation{Materials Science Division, Argonne National Laboratory, Argonne, IL 60439, USA}
	
	\date{\today}
\begin{abstract}
	Non-reciprocity of signal transmission enhances capacity of communication channels and protects transmission quality against possible signal instabilities, thus becoming an important component ensuring  coherent information processing. However, non-reciprocal transmission requires breaking time-reversal symmetry (TRS) which poses challenges of both practical and fundamental character hindering the progress. Here we report a new scheme for achieving broadband non-reciprocity using a specially engineered hybrid microwave cavity. The TRS breaking is realized via strong coherent coupling between a selected chiral mode in the microwave cavity and a single collective spin excitation (magnon) in a ferromagnetic yttrium iron garnet (YIG) sphere. The non-reciprocity in transmission is observed spanning nearly a 0.5 GHz frequency band, which outperforms by two orders of magnitude the previously achieved bandwidths. Our findings open new directions for robust coherent information processing in a broad range of systems in both classical and quantum regimes.
\end{abstract}
	
%\pacs{}
\maketitle

\section{Introduction}

Non-reciprocity protects coherent information processing against noise and backscattering which degrade global coherence~[\onlinecite{Clerk_NComm2018, Metelmann_PRA2018, Alu_NPhys2017}]. Recent implementations of non-reciprocal devices utilized high quality factor resonators~[\onlinecite{Hamann_PRL2018, Lecocq_arxiv2019, Painter_NPhys2017, Ruesink_NC2016, Shen_NPhoton2016, Kim_NPhys2015}]. Most of them suffer a narrow bandwidth and/or the lack of tunability which hindered the wide application of these approaches. An alternative approach employing magnetism~[\onlinecite{Fay_IEEETrans1965}] seems tempting, but recent developments~[\onlinecite{Dietz_PRL2007}, \onlinecite{Dietz_PRL2011}] utilizing ferrites for breaking time-reversal symmetry in microwave cavities brought weak effects due to small dispersive perturbation.

Hybrid systems that couple two or more dynamic excitations promise yet another novel approach for obtaining lacking functionalities. An encouraging example is strongly coupled magnonic systems~[\onlinecite{Huebl_PRL2013, Zhang_PRL2014, Tabuchi_PRL2014, Goryachev_PRL2014, Bai_PRL2015, Tabuchi_Science2015, Kurizki_PNAS2015, Zhang_NComm2015}], where spin waves---the dynamic magnetization excitation in a collective spin ensemble---coherently interact with microwave cavity photons. There, spin waves propagating in cavity and described by quantum electrodynamics (QED) demonstrate a good tunability thus having high potential for quantum applications at macroscopic scale~[\onlinecite{Tabuchi_Science2015}, \onlinecite{Nakamura_SciAdv2017}]. Along the similar path, the non-reciprocity of magnons, which has been observed and utilized for coherent information processing~[\onlinecite{An_NMatt2013, Gladii_PRB2016_nonrecip, Wang_PRL2018Seebeck, Chen_PRB2019_Unidirec}], can be brought into microwave electrodynamics through hybridization. Yet, non-reciprocity has not been reported in hybrid magnon-microwave photon systems until very recently~[\onlinecite{Hu_PRL2019}]. Even there the bandwidth of the non-reciprocal effect remains limited. We address this challenge and develop a chiral-state microwave cavity, where non-reciprocal strong coupling between magnons and microwave photons is achieved, hence enabling large non-reciprocity within a bandwidth that is two orders of magnitude larger than what has been previously reported.

\begin{figure*}[!ht]
	\centering{}
	\includegraphics[width=0.8\linewidth]{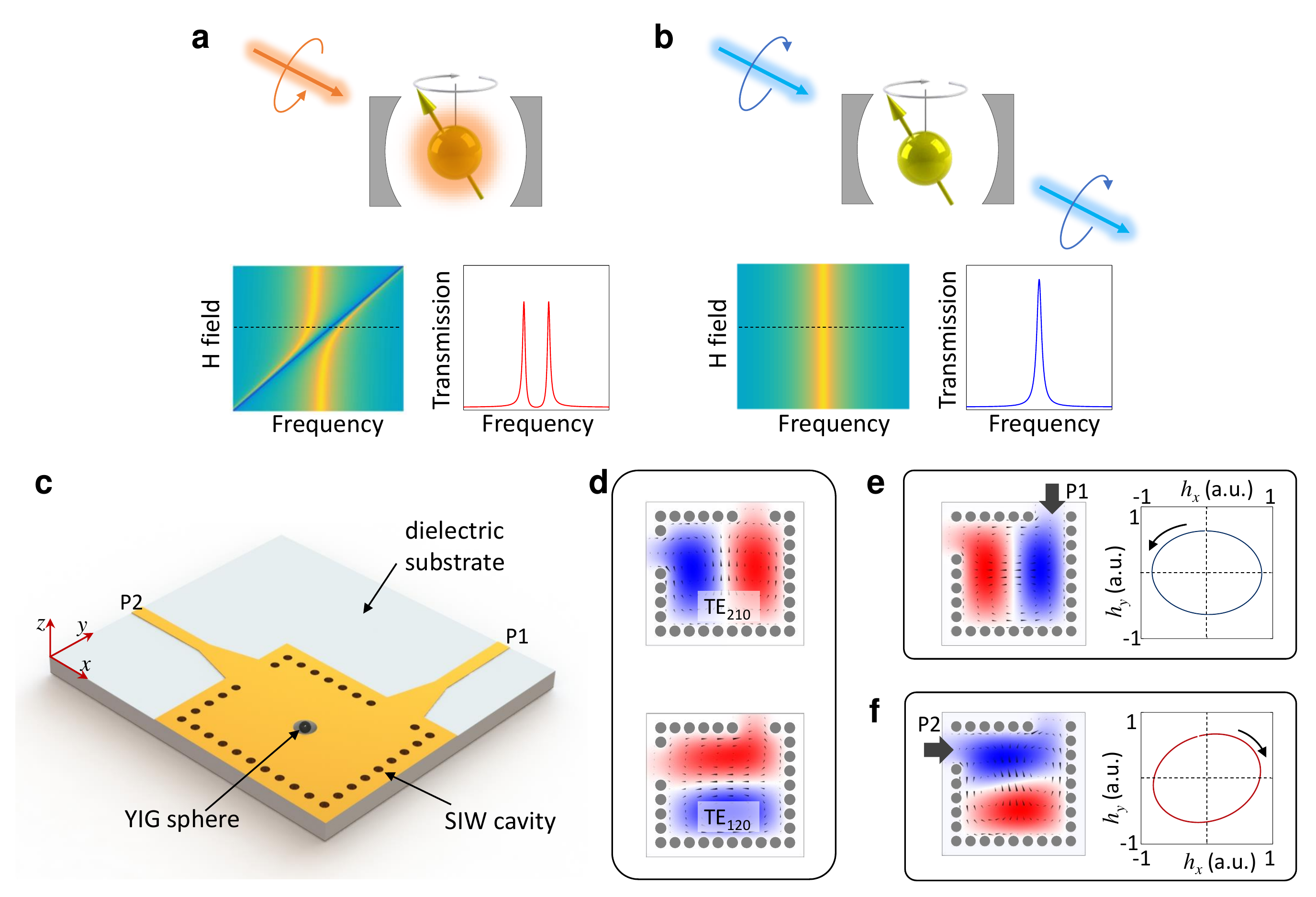}
	\caption{\textbf{Illustration of principles for the non-reciprocal magnon-microwave photon coupling in a microwave cavity hosting a magnon resonator}. \textbf{a}. Magnons and microwave photons with the same chirality couple with each other, resulting in magnon-microwave photon hybridization which appears as the avoided-crossing and normal mode splitting features in the transmission spectrum.  \textbf{b}. Magnons and microwave photons with opposite chiralities do not couple with each other. Microwave photons propagate through the cavity as if there were no magnons present. The transmission spectra are identical to that of a bare cavity. \textbf{c}. Schematic drawing of the substrate-integrated waveguide cavity. Yellow area is the copper layer, which also covers the entire backside of the chip. Gray area is the high-dielectric-constant substrate. The array of black dots are metalized vias that connect the top and bottom copper layer. P1 and P2 are the coupling ports connected to the cavity through tapered microstrips. The hole in the center of the cavity hosts the YIG sphere. \textbf{d}. Simulated cavity field distribution for two orthogonal resonances in a rectangular cavity ($N_x=10$ and $N_y=9$). These two modes have linear microwave magnetic fields at the cavity center along orthogonal directions. Colors and arrows represent the electric and magnetic fields, respectively. \textbf{e}\,\&\,\textbf{f}. Simulated cavity field distribution and field polarization at the center of a square cavity ($N_x=N_y=9$), when the input signal is sent from Port P1 or Port P2, respectively. The excited cavity magnetic fields and electric fields are orthogonal. Near circular polarization is observed, and the chirality is reversed upon swapping the input and output ports. $h_{x}$ ($h_y$) represent $x$ ($y$) components of the microwave magnetic field for the cavity resonance.}
	\label{fig:1}
\end{figure*}

In our approach, a great advance is built on utilizing the chirality of magnons and cavity photons. Magnons being elementary excitation of spin waves inherit the single spin's ability to precess, hence chirality. Therefore, their interaction with chiral microwave photons obeys the selection rule: only magnons and microwave photons with the same chirality interact with each other, while the interaction of magnons and photons with opposite chiralities is forbidden. In the microwave cavity supporting two orthogonal chiral states  (clockwise and counterclockwise circularly polarized photons), the magnons and cavity photons having the same chirality interact with each other and form hybridized modes. Accordingly, avoided-crossing or mode splitting is expected in the transmission spectra (Fig.~\ref{fig:1}a). Once the photon chirality is reversed, the interaction disappears and the photons will pass through the cavity without interactions with magnons, (Fig.~\ref{fig:1}b), hence our cavity will exhibit strong non-reciprocal effects. 

\section{Results}

\subsection{Microwave cavity with chiral states}

%% modes
The cavity is fabricated on a printed circuit board (PCB) following a substrate integrated waveguide (SIW) design (Fig.~\ref{fig:1}c). One-dimensional arrays of metalized vias are arranged on the substrate to form cavity walls. The sub-wavelength via spacing ($1.2$ mm) ensures that the via arrays function as metal boundaries and enclose electromagnetic waves inside the substrate. With a fixed via pitch, the cavity resonance frequency is determined by the number of vias along each direction, $N_x$ and $N_y$. We employ two transverse electric modes, TE$_{120}$ and TE$_{210}$, as illustrated in Fig.~\ref{fig:1}d, whose microwave magnetic fields are the strongest at the cavity center where the magnon resonator is placed. More importantly, when $N_x=N_y$ the cavity possesses a four-fold rotational symmetry. In this case, the two linearly polarized modes are degenerate and can form chiral cavity modes with angular momentum $J_z=\pm 1$.

%% Ports.
In order to excite a particular chiral state, the two linear cavity modes need to be linearly combined with a $\pm\frac{\pi}{2}$ phase difference, which is achieved using a special port design. Two coupling ports of the cavity are engineered by removing two vias near the corners of two neighboring walls, imposing the $\frac{\pi}{2}$ phase offset. This design ensures the degeneracy of the two cavity modes necessary to create a chiral state. As shown in Figs.\,\ref{fig:1}e\,\&\,f, two standing wave cavity modes have a nearly perfect $\frac{\pi}{2}$ phase difference when excited from a single port: the maximum magnetic field of one mode corresponds to the maximum electric field of the other one, and vice versa. Consequently, the ports can excite nearly-circular modes with opposite chiralities. As a result, the selectivity of the interaction between magnons and chiral photon modes described above, imprints non-reciprocity onto the signal transmission in the system. To couple traveling input/output signals to/from the cavity system, a microstrip waveguide is fed to each port with a tapered region for optimized mode matching. Such a SIW design of the cavity circuit system allows for an easy integration with planar structures.

%% YIG sphere
Our magnon resonator is a highly polished single-crystal YIG sphere with the diameter 400\,$\mu$m. The fundamental magnon mode, i.e., the ferromagnetic resonance (FMR), is selected because its uniform mode profile couples most efficiently with the cavity fields. A small hole is drilled at the cavity center to host the YIG sphere, so that maximum magnetic-field mode overlap can be achieved to enhance the magnon-photon coupling. Moreover, the implementation of large-dielectric-constant PCB substrate allows to significantly reduce the volume of the cavity, which further boosts the interaction. The magnon-photon coupling strength can be tuned by moving the YIG sphere along the $z$ direction, see Fig.\,\ref{fig:1}c, using a translational stage. As the sphere is moved deeper into the hole, the coupling strength increases. The magnon frequency is defined as $\omega_{\mathrm m} = \gamma H$, where $\gamma=2\pi\times2.8$\,MHz/Oe is the gyromagnetic ratio, and can be tuned by an external static magnetic field $H$. Since both cavity modes have their microwave magnetic fields along in the $xy$-plane, the external static magnetic field is applied in the out-of-plane direction, i.e., along the $z$-axis, to ensure the simultaneous interaction between the magnon mode and both cavity modes. 

\begin{figure*}[!t]
	\centering{}
	\includegraphics[width=1\linewidth]{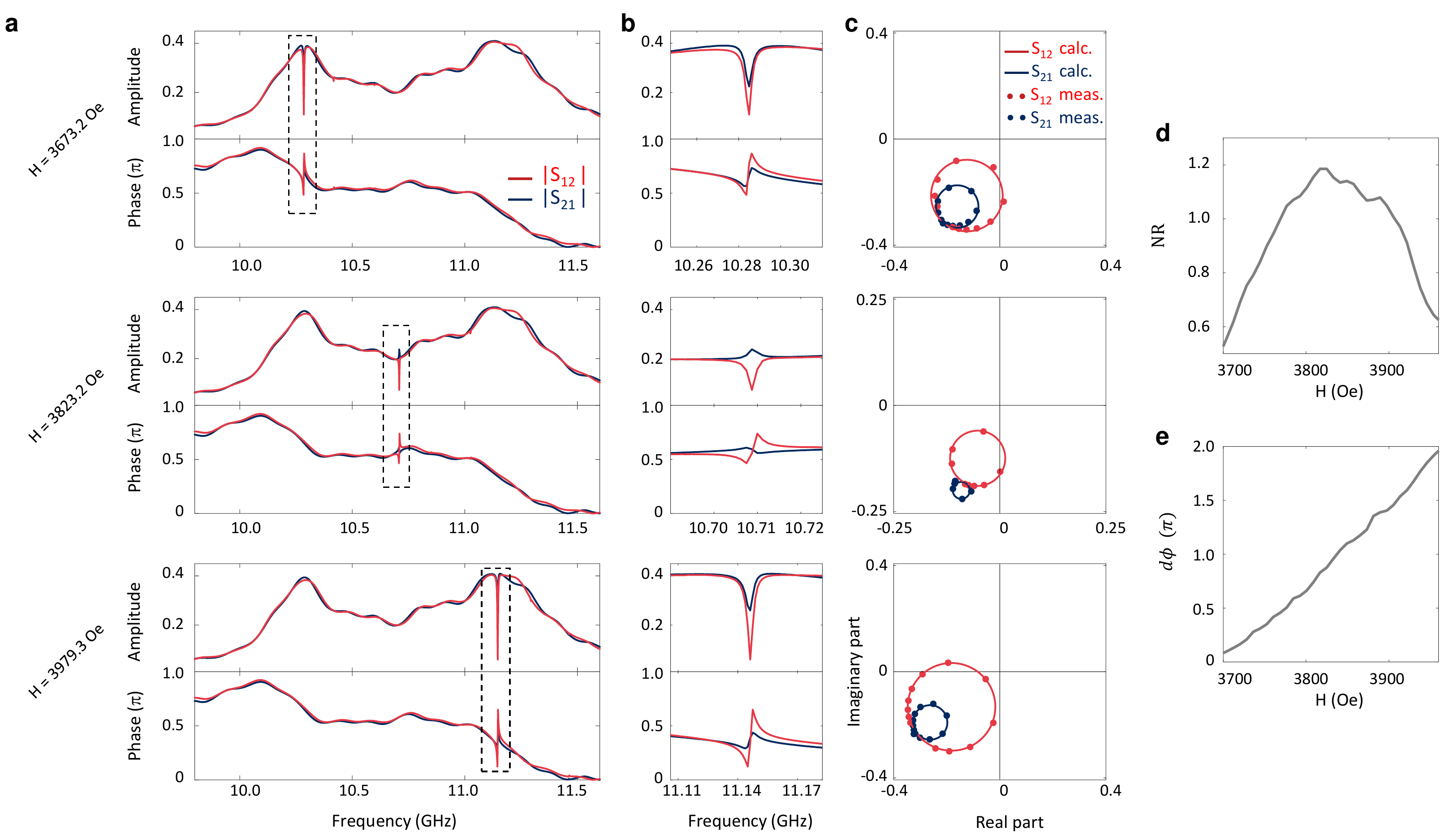}
	\caption{\textbf{Non-reciprocal magnon-microwave photon interaction}. \textbf{a}. Measured amplitude and phase of $S_{12}$ (red) and $S_{21}$ (blue) spectra from the rectangular cavity ($N_x=10$ and $N_y=9$) at three selected bias magnetic fields: $H=$ 3673.2 Oe, 3823.2 Oe, and 3979.3 Oe, respectively. \textbf{b}. Zoomed-in spectra of the magnon resonances (as indicated by the dashed boxes in \textbf{a}) at the three selected bias conditions. \textbf{c}. Polar plot of the magnon resonances. Dots are measurement results, and solid lines are theoretical calculations. \textbf{d}\,\&\,\textbf{e}. Summery of the extracted non-reciprocity $NR$ and phase differences $d\phi$. Maximum non-reciprocity and a $\pi$ phase difference is observed around $H=3823$ Oe.}
	\label{fig:2}
\end{figure*}

\subsection{Theoretical model}

In the most general case, the two cavity modes under consideration can be non-degenerate and have different resonant frequencies and coupling strengths with the magnon mode. The system Hamiltonian is

\begin{equation}
H=\hbar\begin{pmatrix}
	\omega_{\mathrm a}-i\frac{\kappa_{\mathrm a}}{2} & 0 & -g_{\mathrm a}\\
	0 & \omega_{\mathrm b}-i\frac{\kappa_{\mathrm b}}{2} & -g_{\mathrm b}\\
	-g_{\mathrm a} & -g_{\mathrm b} & \omega_{\mathrm m}-i\frac{\kappa_{\mathrm m}}{2}
	\end{pmatrix}\,,
	\label{eq:Hamiltonian}
\end{equation}
\noindent where $\omega_x$ and $\kappa_x$ ($x=a,b,m$) are the resonant frequencies and total dissipation rates of the two microwave modes, $a$ and $b$, and the magnon mode $m$, respectively, and $g_\mathrm{a,b}$ are the coupling strength between the magnon mode and each microwave mode.
%The non-reciprocal effects depend on not only the intrinsic properties of the system (described by the Hamiltonian in Eq.\,\ref{eq:Hamiltonian}), but also how microwave photons are coupled in and extracted out from the ports.

The equation of motion for the composite intra-cavity field, $\textbf{u}=(a,b,m)^T$, can be written as 
\begin{equation}
\frac{d}{dt}\textbf{u}=\frac{H}{i\hbar}\textbf{u}+B\textbf{s}_\mathrm{in}\,,
\end{equation}
where $\mathbf{s}_\mathrm{in}=(s_\mathrm{in1},s_\mathrm{in2})^T$ describes input fields from ports P1 and P2, and $B$ is a 3-by-2 matrix of coupling constants of the ports, given by
\begin{equation}
B=\begin{pmatrix}
	\sqrt{\kappa_\mathrm{ae1}} & \sqrt{\kappa_\mathrm{ae2}}e^{i\alpha} \\
	\sqrt{\kappa_\mathrm{be1}}e^{i\beta} & \sqrt{\kappa_\mathrm{be2}}e^{i(\alpha+\beta+\pi)} \\ 0 & 0
	\end{pmatrix}\,.
\end{equation}
Here, $\kappa_\mathrm{ae(1,2)}$ and $\kappa_\mathrm{be(1,2)}$ are the external coupling rates between the microwave modes and the two ports, $\alpha$ is the relative phase difference of one cavity mode when it is excited from the two different ports, while $\beta$ describes the phase difference between the two modes when they are excited from the same port. Note that the last row in $B$ is zero because we assume no direct excitation of magnons from the ports. 

In the general non-degenerate case, modes $a$ and $b$ indices correspond to the two standing-wave modes in our cavity system as shown in Fig.\,\ref{fig:1}d. Since for standing waves the cavity field at different position is always in phase, we have $\alpha\approx0$. As discussed above, in our configuration, $\beta\approx\frac{\pi}{2}$, which leads to circular states inside the cavity when the two linearly polarized standing-wave modes are similar in amplitude. Using the input-output relation, one finds the scattering matrix of the system, see Appendix for details.

\begin{figure*}[ht]
	\centering{}
	\includegraphics[width=1\linewidth]{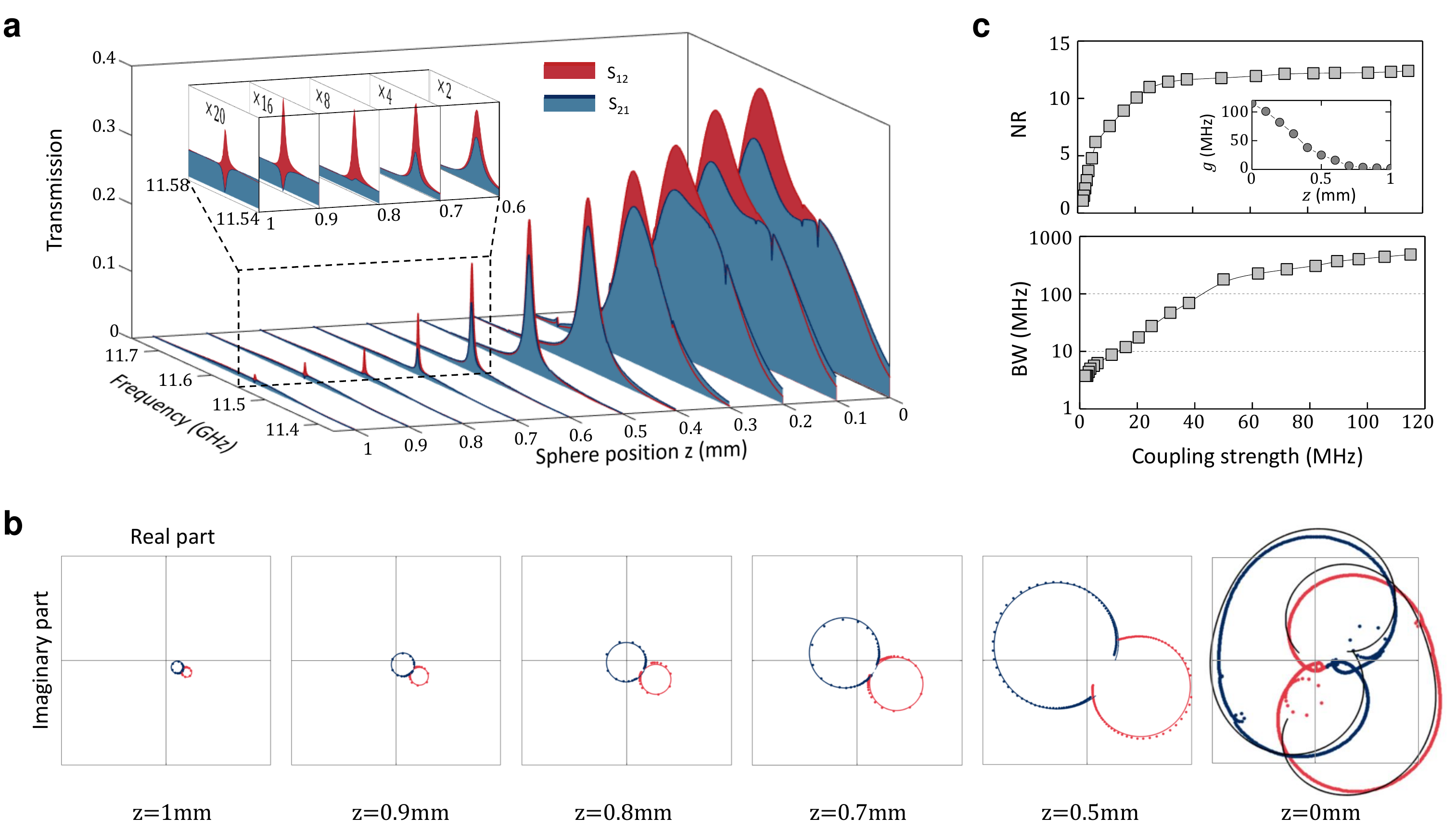} 
	\caption{\textbf{Coupling strength-dependent non-reciprocal transmissions.} \textbf{a}. Transmission spectra at various sphere positions. $z=0$ mm corresponds to the cavity center. When the sphere is moved away from the cavity center, the overall transmission level decreases. \textbf{b}. Polar plot of the cavity transmission at selected YIG sphere positions. The non-reciprocity is evidently shown by the opposite phases of the magnon resonance circles measured in $S_{21}$ and $S_{12}$. Blue (red) lines and dots correspond to the calculated and measured $S_{21}$ ($S_{12}$), respectively. \textbf{c}. Non-reciprocity (NR, upper panel) and bandwidth (BW, lower panel) as a function of the coupling strength, both of which show an increasing trend with coupling strength. Inset: coupling strength as a function of YIG sphere position.}
	\label{fig:3}
\end{figure*}

\subsection{Non-reciprocity and photon chirality}

We begin with the non-reciprocal effect in a non-degenerate cavity in the set up where microwave modes have different frequencies.
%polarizations are separated and can be individually controlled. 
When $N_x=10$ and $N_y=9$, the two orthogonal cavity modes resonate at $\frac{\omega_{\mathrm a}}{2\pi}=10.285\,\textrm{GHz}$ and $\frac{\omega_{\mathrm b}}{2\pi}=11.142\,\textrm{GHz}$. The transmission spectra recorded using a Vector Network Analyzer (VNA) are plotted in Fig.~\ref{fig:2}a. The two broad transmission peaks correspond to TE$_{120}$ and TE$_{210}$ modes, respectively. The magnon resonances show up as the sharp features in the dashed boxes. To investigate the influence of the photon chirality on the magnon-photon interaction, we tune the magnon resonance by changing the bias magnetic field. Three conditions are plotted: $H=$ 3673.2 Oe, 3823.2 Oe, and 3979.3 Oe, corresponding to $\frac{\omega_{\mathrm m}}{2\pi}=$ 10.285 GHz, 10.705 GHz, and 11.142 GHz, respectively. Zoomed-in plots are shown in Fig.\,\ref{fig:2}b to reveal details. The polarization of cavity photons is determined by the relative strength of the two linearly polarized modes. As a result, the circularity of the photon polarization varies at different frequencies, leading to different non-reciprocal effects.

In the middle of the two cavity resonances, photons exhibit nearly perfect circular polarization, since the contributions from the two linear modes are close to each other. When the magnon mode is properly tuned, non-reciprocity is observed and the magnon resonance manifests as a peak in $|S_{21}|$ and as a dip in $|S_{12}|$ (see the middle panel in Fig.\,\ref{fig:2}b). The phases of $S_{12}$ and $S_{21}$ also show a distinct changes in the opposite directions as seen in Fig.~\ref{fig:2}c when plotting in the complex plane.

On the contrary, the non-reciprocity becomes much weaker when the magnon is in the resonance with the cavity mode at $H=$ 3673.2\,Oe or 3979.3 \,Oe, where the cavity photons are dominated by a single linear polarization. Under these conditions, the magnon resonance appears as a dip for both $S_{21}$ and $S_{12}$, and their phase changes are alike, which in the ideal system would cancel the non-reciprocity. It still remains finite due to a finite cavity linewidths, providing a non-zero contribution from the orthogonal polarization. In the polar plot, the two magnon resonances are in-phase with only slight  differences in the amplitude.

The behaviour of the non-reciprocity is quantitatively summarized in Fig.\,\ref{fig:2}d, where it is defined as $\mathrm{NR}=\left|\left({S_{12}-S_{21}}\right)/{S_0}\right|$, with $S_0$ being the cavity transmission in the absence of the magnon mode. Note that $S_{12}$, $S_{21}$, and $S_0$ are all complex numbers, and, therefore, not only the amplitude but also the phase is included in the non-reciprocity. It is evident that non-reciprocity reaches its maximum at $H=3823.2$ \,Oe. Figure\,\ref{fig:2}e shows the relative phase, $d\phi=\angle{(S_{12}-S_0)}-\angle{(S_{21}-S_0)}$, as a function of the bias magnetic field. The relative phase has a nearly linear dependence on the bias field, changing from 0 to $2\pi$ as the magnon frequency moves from one cavity mode to the other, with the maximum phase difference of $\pi$ reached half way through.

\begin{figure*}[!ht]
	\centering{}
	\includegraphics[width=0.8\linewidth]{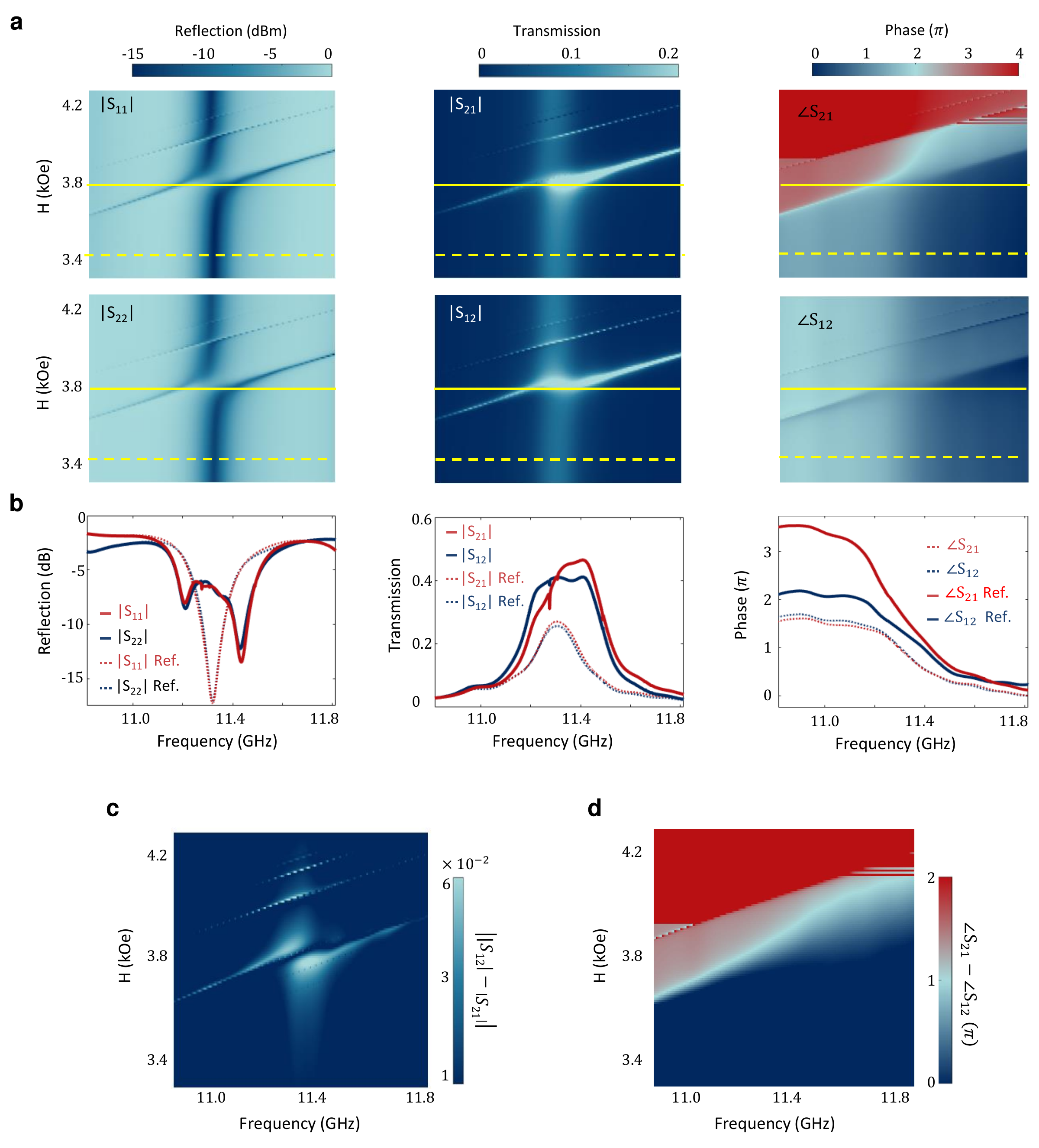}
	\caption{\textbf{Non-reciprocal magnon-photon strong coupling}. \textbf{a}. Measured reflection and transmission spectra at various bias magnetic fields in a square cavity ($N_x=N_y=9$). The two degenerate cavity resonances show up in the spectra as a single resonance dip/peak at 11.32 GHz. The reflection spectra from both ports show similar avoided crossing features at $H=3800$ Oe, indicating the strong coupling condition between magnons and microwave cavity photons. The small avoid crossings at higher bias fields are the results of high-order magnon modes. The transmission spectra exhibit distinctly different features in both amplitude and phase for the two opposite transmission directions ($S_{12}$ v.s. $S_{21}$). \textbf{b}. Solid lines: reflection and transmission spectra for on-resonance conditions, as indicated by the yellow solid lines in \textbf{a}. Dotted lines: reference signals when magnons are detuned far off resonance, as indicated by the yellow dashed lines in \textbf{a}. Compared with the off-resonance conditions, the on-resonance reflections clearly show the normal mode splitting. When magnon is off-resonance, the transmissions for both signal directions ($S_{12}$ and $S_{21}$) are identical, but when magnon is on-resonance, they show unambiguous differences in both the amplitude and phase. \textbf{c} and \textbf{d}. Differences between the amplitude and phase of $S_{12}$ and $S_{21}$ as a function of frequency and bias magnetic field, respectively. The differences reach maximum when the magnon is tuned on-resonance with the cavity resonances.} 
	\label{fig:4}
\end{figure*}

\subsection{Non-reciprocal strong coupling}

To further enhance the non-reciprocal effect, we design a square cavity ($N_x=N_y=9$) with degenerate TE$_{210}$ and TE$_{120}$ modes ($\frac{\omega_{\mathrm a}}{2\pi}=\frac{\omega_{\mathrm b}}{2\pi}=11.30$ GHz). Because of their identical frequency, both modes can be simultaneously excited by either of the two coupling ports. Due to the symmetric cavity geometry, we have $\kappa_{\mathrm a}=\kappa_{\mathrm b}$ and $g_{\mathrm a}=g_{\mathrm b}$. As a result, the two linear modes form nearly perfect circular polarizations (Figs.\,\ref{fig:1}e\,\&\,f), with chirality determined by the excitation port. In a circular mode basis, the system's Hamiltonian can be rewritten as
\begin{equation}
H_\mathrm{cir}=\hbar\left(\begin{array}{ccc}
\omega_\mathrm{a}-i\frac{\kappa_\mathrm{a}}{2} & 0 & -\sqrt{2}g_\mathrm{a}\\
0 & \omega_\mathrm{a}-i\frac{\kappa_\mathrm{a}}{2} & 0\\
-\sqrt{2}g_\mathrm{a} & 0 & \omega_\mathrm{m}-i\frac{\kappa_\mathrm{m}}{2}
\end{array}\right)\,,
\end{equation}
\noindent demonstrating the selection rule for polarization: only one circularly polarized cavity mode interacts with the magnon mode, while the other cavity mode is decoupled.

The combination of the polarization selectivity and the port-dependent chirality enables the on-resonance non-reciprocal magnon-photon coupling. Based on our model, the transmissions for both directions are calculated as:
\begin{eqnarray}
S_{12,21}=\pm\frac{ig_{\mathrm a}^2\kappa_\mathrm{a}}{(\frac{\kappa_{\mathrm a}}{2}-i\Delta_{\mathrm a})\left[2g_{\mathrm a}^2+(\frac{\kappa_{\mathrm a}}{2}-i\Delta_{\mathrm a})(\frac{\kappa_{\mathrm m}}{2}-i\Delta_{\mathrm m})\right]},
\label{Sij}
\end{eqnarray}
\noindent where $\Delta_\mathrm{a,m} \equiv \omega-\omega_\mathrm{a,m}$.

The cavity transmission is measured at different positions $z$ of the YIG sphere, as shown in Fig.\,\ref{fig:3}. The calculated results are in an excellent agreement with the complex values of the measured spectra over the range of positions $z$. At large $z$, the sphere is far away from the center, $z = 0$, and the overlap between the cavity photon modes and the magnon mode is small, leading to a small magnon-photon coupling strength $g$. The non-reciprocity similar to what seen in the non-degenerate case is observed: the magnon resonance results in a peak in $S_{12}$ and in a dip in $S_{21}$, see Fig.\,\ref{fig:3}a. Because of the small coupling strength, the peaks and dips are relatively narrow. 

As the sphere moves closer to the center, the coupling strength grows with increasing overlap between the magnon and photon modes. This modifies the resonance line shapes in transmission spectra. At $z=0.9$\,mm, the dip in $S_{21}$ reaches zero, leading to the infinitely large isolation ratio between the forward and backward propagation. At $z=0.8$ mm, the dip almost disappears from $S_{21}$, causing magnon invisibility~[\onlinecite{Hu_PRL2019}]. As $z$ is decreased further, the dip in $S_{21}$ turns into a peak. On the contrary, the magnon resonance always appears as a peak in $S_{12}$. The magnon-photon interaction increases significantly with the coupling strength $g$ and becomes dominant over the overall device transmission when the YIG sphere is close to the cavity center. At the same time, the linewidth of the transmission peak also becomes significantly larger.

The transmission amplitude spectra show that at smaller $z$, i.e., larger $g$, the difference between $S_{12}$ and $S_{21}$ becomes smaller. Nonetheless, the polar plots in Fig.\,\ref{fig:3}b reveal that the magnon resonances measured in the two opposite directions always have a $\pi$ phase difference, regardless of the coupling strength. At larger $z$ (smaller $g$) values, the magnon resonance is comparable with the crosstalk signal in amplitude, and, therefore, their interference effect is large, giving rise to highly direction-dependent transmission amplitude. At smaller $z$ (larger $g$) values, the magnon resonance is dominant, while the crosstalk signal is negligible, resulting in weaker interference effects in the transmission amplitude. Both amplitude and phase play an important role for coherent information processing and, therefore, both are essential for quantifying non-reciprocity. The increasing coupling strength enhances non-reciprocity as shown in Fig.\,\ref{fig:3}c.

In addition to enhancing the non-reciprocity, the larger magnon-photon coupling strength is crucial for increasing the range of frequencies at which strong non-reciprocity is observed. In our system, although the measured non-reciprocity stays at the relatively unchanging level when $\frac{g}{2\pi} > 30$ MHz, the bandwidth of the non-reciprocal interaction keeps increasing with the coupling strength, as plotted in the bottom panel of Fig.\,\ref{fig:3}c. When the coupling strength is small, Eq.\,(\ref{Sij}) indicates that the bandwidth of the non-reciprocity is limited by the magnon linewidth ($\mathrm{BW}\approx\frac{\kappa_\mathrm{m}}{2}$). In the opposite limit of the large coupling strength ($g\gg\frac{\kappa_{\mathrm{m}}}{2},\frac{\kappa_{\mathrm a}}{2}$), the non-reciprocity bandwidth becomes $\mathrm{BW}\approx\frac{\kappa_{\mathrm a}}{2}$, according to Eq.\,(\ref{Sij}), indicating that the magnon-induced non-reciprocity can take place throughout the whole cavity linewidth. Interestingly, when the coupling strength $g$ is comparable with $\frac{\kappa_{\mathrm a}}{2}$, the non-reciprocity bandwidth can be even larger because of the complicated transmission lineshape. In our experiments, a maximum bandwidth of 482 MHz is recorded, which is more than two orders of magnitude larger than in the non-degenerate case or previous demonstrations which are limited by the magnon linewidth.

Importantly, the interacting magnon-photon system in our experiment achieves a strong coupling regime. In particular, when the sphere is moved to the center, $z=0$, the coupling strength of 115 MHz is reached, which exceeds the dissipation rate of individual cavity mode (110 MHz) and the magnon mode (1 MHz). The strong coupling regime is confirmed by the avoided crossing observed in the cavity reflection coefficients, shown in Fig.~\ref{fig:4}a, as well as by the clearly resolved normal mode splitting in Fig.~\ref{fig:4}b. The transmission spectra, Fig.~\ref{fig:4}a (middle and right column), reveal the non-reciprocity. When the magnon frequencies are tuned away from the cavity resonance (e.g., at $H=3.5$ kOe), the $S_{12}$ and $S_{21}$ spectra are almost identical, as indicated by the dotted lines in Fig.~\ref{fig:4}b. However, when the magnon mode is on resonance with the cavity photon modes, the $S_{12}$ and $S_{21}$ spectra become distinctly different. The non-reciprocity is even more prominent in the phase map. At $H=3.5$ kOe when magnon is off resonance, the phase response of $S_{12}$ and $S_{21}$ are nearly identical (dashed lines), with a 2$\pi$-phase change across the cavity resonance. When the magnon is in resonance, the phase response in $S_{12}$ is only slightly different from the off-resonant one, while in $S_{21}$ it changes drastically, with a $4\pi$ phase change across the resonance frequency. This phase behavior is visualized separately in Figs.~\ref{fig:4}c\,\&\,d, clearly illustrating the non-reciprocity. Note there are two high-order magnon modes visible at higher bias fields, which appear as two narrow lines in the amplitude but induce large changes in phase.

\section{Discussion}

The observed chirality-based non-reciprocal transmission based on the strong magnon-photon coupling significantly increased the operating frequency range by two orders of magnitude opening route for the broadband non-reciprocal coherent information processing. Furthermore, the original construction of our cavity provided exceptional controllability where the bandwidth is tuned by the position of the YIG sphere inside the cavity and the operation frequency is tuned by the bias magnetic field. Thus, our findings point out a direction for a novel emergent class of non-reciprocal devices for coherent information transmission. The next forthcoming step is its application in the quantum regime where unidirectional signal propagation is critical for mitigating detrimental effects of noise and fluctuations.

\begin{acknowledgments} The work at the Center for Nanoscale Materials, a U.S. Department of Energy Office of Science User Facility, was supported by the U.S. Department of Energy, Office of Science, under Contract No. DE-AC02-06CH11357, the work at MSD was supported by the U.S. Department of Energy, Office of Science, Basic Energy Sciences, Materials Sciences and Engineering Division. We thank the Chicago Quantum Exchange for facilitating this research.
\end{acknowledgments}

%%%%%%%%%%%%%%%%%%%%%%%%%%%%%%%%%%%%%%%%%%%%%%%%%%%%%%%%%%%%%%%%%%%%%%%%%%%%%%
%%%%%%%%%%%%%%%%%%%%%%%%%%%%%%%%%%%%%%%%%%%%%%%%%%%%%%%%%%%%%%%%%%%%%%%%%%%%%%
%%%%%%%%%%%%%%%%%%%%%%%%%%%%%%%%%%%%%%%%%%%%%%%%%%%%%%%%%%%%%%%%%%%%%%%%%%%%%%
%%%%%%%%%%%%%%%%%%%%%%%%%%%%%%%%%%%%%%%%%%%%%%%%%%%%%%%%%%%%%%%%%%%%%%%%%%%%%%
%%%%%%%%%%%%%%%%%%%%%%%%%%%%%%%%%%%%%%%%%%%%%%%%%%%%%%%%%%%%%%%%%%%%%%%%%%%%%%
%%%%%%%%%%%%%%%%%%                               %%%%%%%%%%%%%%%%%%%%%%%%%%%%%
%%%%%%%%%%%%%%%%%%           Appendix            %%%%%%%%%%%%%%%%%%%%%%%%%%%%%
%%%%%%%%%%%%%%%%%%                               %%%%%%%%%%%%%%%%%%%%%%%%%%%%%
%%%%%%%%%%%%%%%%%%%%%%%%%%%%%%%%%%%%%%%%%%%%%%%%%%%%%%%%%%%%%%%%%%%%%%%%%%%%%%
%%%%%%%%%%%%%%%%%%%%%%%%%%%%%%%%%%%%%%%%%%%%%%%%%%%%%%%%%%%%%%%%%%%%%%%%%%%%%%
%%%%%%%%%%%%%%%%%%%%%%%%%%%%%%%%%%%%%%%%%%%%%%%%%%%%%%%%%%%%%%%%%%%%%%%%%%%%%%
%%%%%%%%%%%%%%%%%%%%%%%%%%%%%%%%%%%%%%%%%%%%%%%%%%%%%%%%%%%%%%%%%%%%%%%%%%%%%%
%%%%%%%%%%%%%%%%%%%%%%%%%%%%%%%%%%%%%%%%%%%%%%%%%%%%%%%%%%%%%%%%%%%%%%%%%%%%%%

\appendix

\section{Cavity design and preparation}

\subsection{Symmetry of the cavity modes}~~\\
In the degenerate case, the square-shaped microwave cavity has $D_{4h}$ point-group symmetry, namely, 4-fold rotational symmetry about $z$ axis, 2-fold rotational symmetry about $x$ and $y$ axes, and reflection symmetry with respect to the $xy$-plane. The eigen-wavefunctions of the electromagnetic field in this cavity necessarily factorize according to
\begin{equation}
	\begin{pmatrix}
		\boldsymbol{\mathcal{E}} \\
		\boldsymbol{\mathcal{H}}
	\end{pmatrix}
	\!=\!
	\begin{pmatrix}
		\boldsymbol{E}_\nu(r,\phi,z) \\
		\boldsymbol{H}_\nu(r,\phi,z)
	\end{pmatrix}
	\Ee^{\Ii \nu \phi}\,,\,\, \text{with}\,\,
	\begin{pmatrix}
		\boldsymbol{E}_\nu\left(r,\phi\!+\!\frac{\pi}{2},z\right) \\
		\boldsymbol{H}_\nu\left(r,\phi\!+\!\frac{\pi}{2},z\right)
	\end{pmatrix}
	\!=\!
	\begin{pmatrix}
		\boldsymbol{E}_\nu(r,\phi,z) \\
		\boldsymbol{H}_\nu(r,\phi,z)
	\end{pmatrix},
\end{equation}
where $\nu=0,\pm1,2$ is the quasi-angular momentum $J_z$ (we omit the Plank constant $\hbar$ here). The $J_z=0$ mode is a monopolar mode. The $J_z = \pm1$ are dipolar modes and transform to each other under time reversal. The $J_z = 2$ mode is a quadruple mode. In principle, we can define a $J_z=-2$ mode, which however is identical to the $J_z=+2$ mode, since they both lie at the Brillouin zone boundary of angular momentum. In this system, only the $J_z=+1$ and $J_z=-1$ modes have definitive, right and left chirality with respect to the $z$-axis. They degenerate with each other before time reversal breaking.

\begin{figure}[!b]
	\centering{}
	\includegraphics[width=\linewidth]{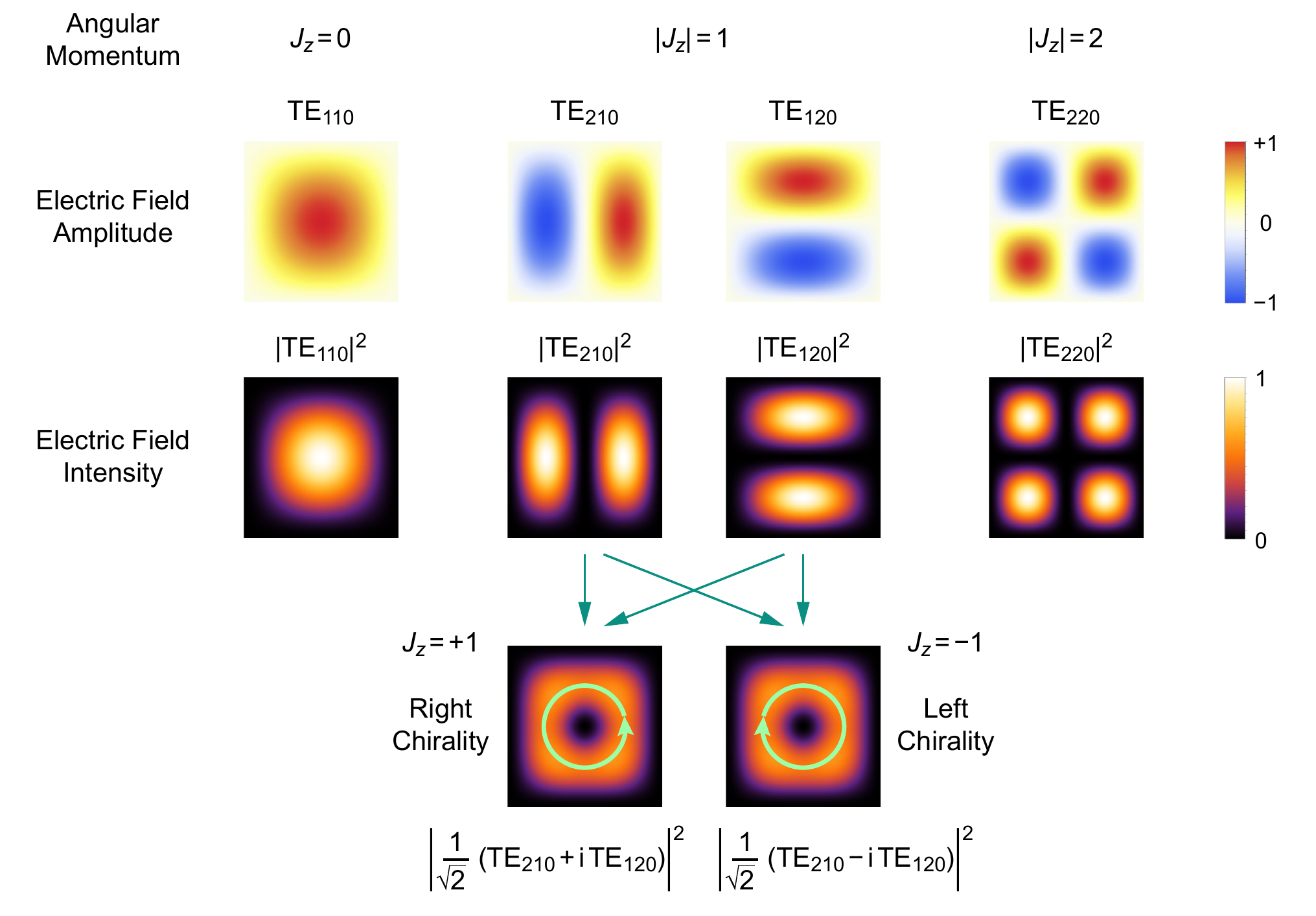}
	\caption{\textbf{Mode analysis}. Classification of cavity photon modes according to the quasi-angular momentum, and the formation of right- and left-chiral modes by linear combinations of TE$_{210}$ and TE$_{120}$ modes.}
	\label{figSM:figSM_AngularMomentum}
\end{figure}

Our main interest lies in the low-energy TE$_{mn0}$ like modes, in which case $\mathcal{E}_z$, $\mathcal{H}_x$ and $\mathcal{H}_y$ are the only nonzero field components and are invariant along $z$. Furthermore, $\mathcal{E}_z$ is the governing component, since $\mathcal{H}_x$ and $\mathcal{H}_y$ can be deduced from $\mathcal{E}_z$ via derivatives. Hence we can write
\begin{equation}
	\mathcal{E}_z = E_{z,\nu}(r,\phi) \Ee^{\Ii \nu \phi}\quad \text{with}\quad E_{z,\nu} \left( r,\phi+\frac{\pi}{2} \right) = E_\nu\left( r,\phi\right).
\end{equation}
The lowest-energy $\nu=0$ and $\nu=2$ mode, respectively, is the conventional TE$_{110}$ mode and TE$_{220}$ mode (see Fig.\,\ref{figSM:figSM_AngularMomentum}). The lowest-energy $\nu=\pm1$ modes are linear combinations of the conventional TE$_{210}$ mode and TE$_{120}$ mode with $\pm\frac{\pi}{2}$ phase difference,
\begin{equation}
	|\nu=\pm1\rangle = \frac{1}{\sqrt{2}} \left(|\text{TE}_{210}\rangle \pm \mathrm{i} |\text{TE}_{120}\rangle\right).
\end{equation}
The angular momentum $J_z=\nu=\pm1$ corresponds to the right- and left-chirality. These two modes have the strongest magnetic field at the center $r=0$. This is crucial for realizing strong magnon-photon coupling. The magnetic field rotates in the counterclockwise and clockwise direction; only one of them can couple with the magnon mode of the same chirality. 

Figure\,\ref{figSM:figSM_AngularMomentum} displays all the profiles of field amplitude $E_z$ and intensity $|E_z|^2$. 

\subsection{Device design and fabrication}

Substrate integrated waveguide (SIW) is a unique technique typically used in integrated microwave circuits which can provide improved mode confinement and loss. Because of the planar nature, they can be easily integrated with printed circuit boards (PCBs). In a SIW, the electromagnetic waves are confined inside the substrate by the top and bottom metal layers. The lateral confinement is achieved by arrays of metal plated via. When the via spacing is smaller than half of the wavelength, their arrays function as metal walls. 

The cavity design is carried out using finite element simulations. Because the modes we choose ($\mathrm{TE_{120}}$ and $\mathrm{TE}_{210}$) have uniform mode distributions along $z$ direction, the structure can be simplified to two-dimensional (2D) for the simulation to speed up the simulation and improve the mesh density. Both eigenmode and driven response are simulated. Eigenmode simulation allows us to find out the resonance frequency of the eigenmodes (standing waves), while driven response allows direct observation of the circular excitation in the cavity.

\begin{figure}[!t]
	\centering{}
	\includegraphics[width=\linewidth]{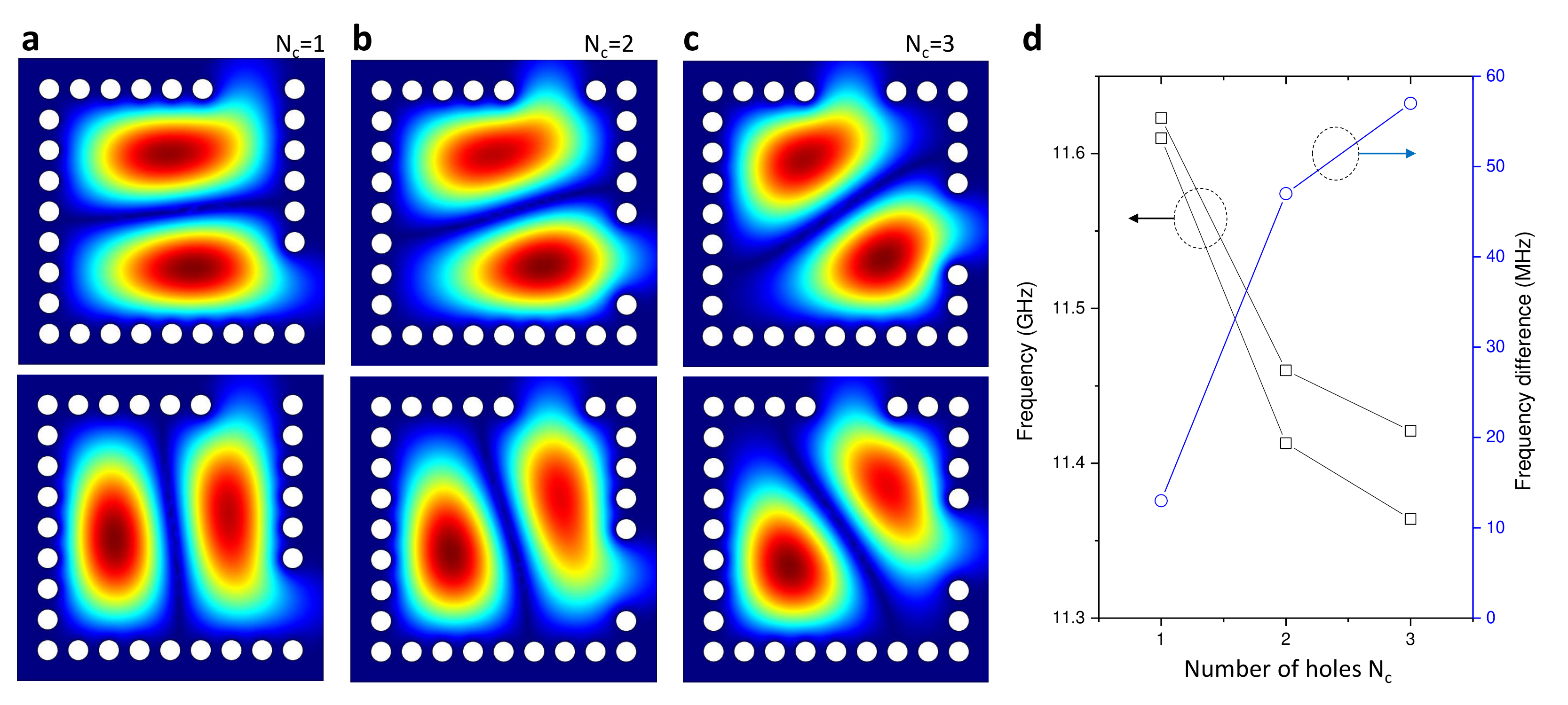}
	\caption{\textbf{Port configuration and mode degeneracy}. \textbf{a}--\textbf{c}. Simulated eigenmode distributions with different port configurations. Top and bottom rows correspond to the two orthogonal modes, respectively. \textbf{d}. Summary of the eigenmode frequencies of the two modes and their differences.}
	\label{figSM:portConfig}
\end{figure}

In our device, a Rogers TMM10i substrate with a thickness of 1.27 mm is used. Because of the large dielectric constant of the substrate (10), the cavity volume is significantly reduced compared with air-filled metal cavities. For further reducing the cavity volume and enhancing the magnon-photon coupling, relatively high resonance frequencies in the range of 10--12 GHz are chosen. The corresponding wavelength inside the dielectric is around 10 mm. As a comparison via spacing of 1.2 mm is used in our device, which is much smaller than half of the wavelength, ensuring the good confinement of the via arrays. In our experiments, the number of holes along each direction is 9 or 10, corresponding to a lateral dimension of 10.8 or 12 mm. The via diameter is 0.8 mm, which is not very critical in the design. The vias are effectively plated by filling them with silver epoxy, which electrically connects the top and bottom copper planes. The relatively high loss of the silver epoxy limits the quality factor of our cavity. 

At the center of the cavity, a hole with a diameter of 1 mm is drill to host the YIG sphere which has a diameter of 400 $\mu$m. The sphere is highly polished to ensure good device performance. It is glued on a ceramic rod as a mechanical support for handling convenience. 

\subsection{Port configuration}

According to the main text, the port configuration plays an critical rule in achieving the non-reciprocity. Therefore, efforts are taken to optimize the port design. The ports for a SIW cavity are usually formed by missing vias, where the confined electromagnetic waves can leak in and out. In our design, because of the relatively large cavity dissipation rage (around 100 MHz), the port coupling rate needs to be high to obtain good extinction ratio when measuring the cavity reflection or transmission. As a result, two adjacent vias (instead of one) are removed to form a coupling port.

During the eigenmode simulation, the ports are included in the geometry to evaluate their effects on the eigenfrequency. Based on the analysis in the main text, it is critical to have the two ports positioned with a $\frac{\pi}{2}$ phase difference. While maintaining such a phase relation, different port positions are tested, and the results are summarized in Fig.\,\ref{figSM:portConfig}. The relative port position is indicated by the number of holes $N_c$ between the port and the corner. As the port moves away from the corner ($N_c$ increases), the eigenfrequencies of both modes reduce. However, this also causes larger mode non-degeneracy. When $N_c=3$, the frequency difference between the two modes is 58 MHz, while it is only 10 MHz when $N_c=1$. Therefore, in our experiments, the $N_c=1$ design is selected when measuring the non-reciprocity to ensure good degeneracy.

\begin{figure}[!t]
	\centering{}
	\includegraphics[width=0.8\linewidth]{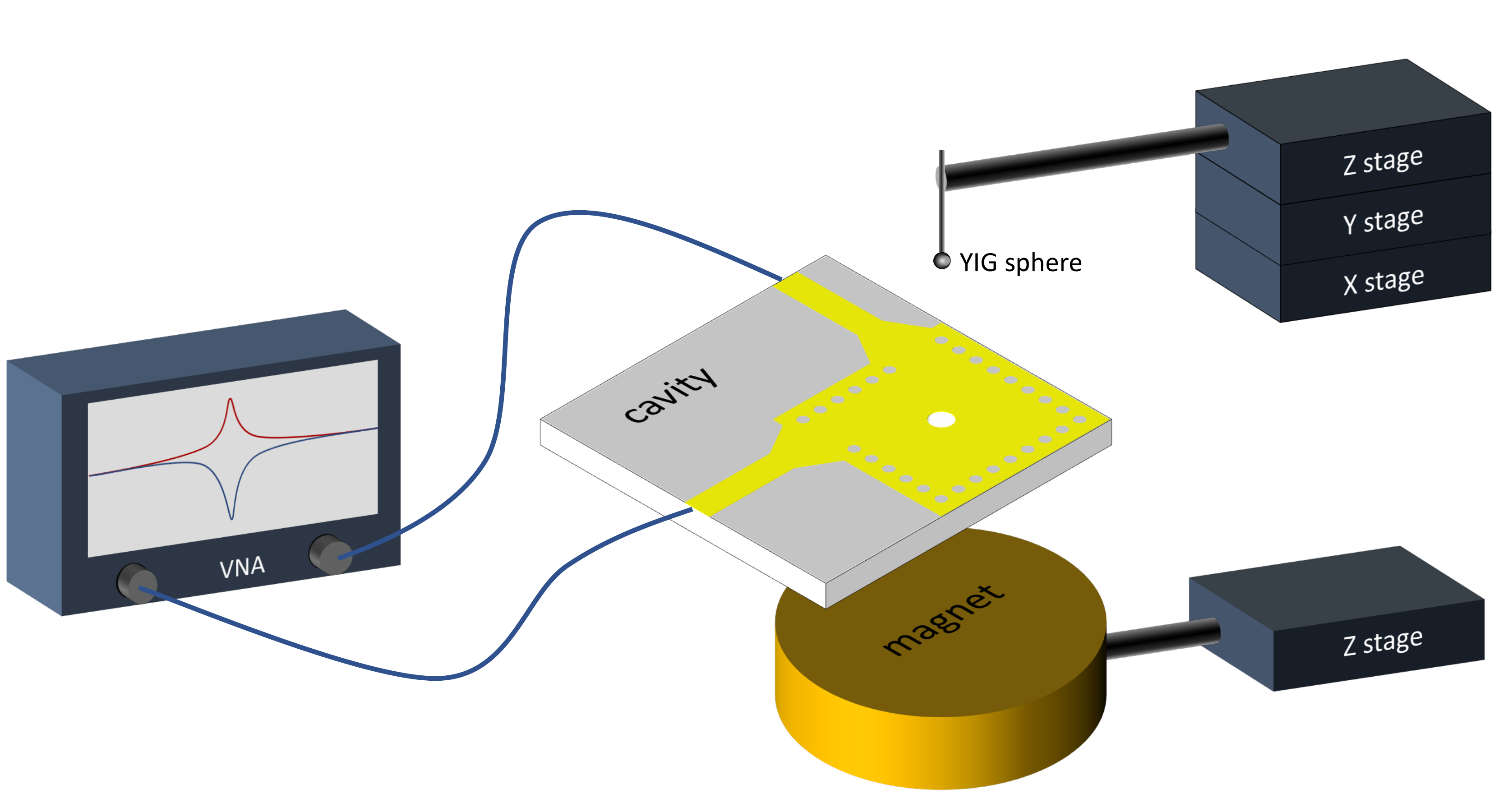}
	\caption{{Schematics of the measurement setup.}}
	\label{figSM:setup}
\end{figure}

\subsection{Device characterization}

The device is characterized using a vector network analyzer (VNA), as plotted in Fig.\,\ref{figSM:setup}. Both reflection and transmission measurements are performed and all the four S parameters are obtained: $S_{11}$, $S_{22}$, $S_{21}$, $S_{12}$. The YIG sphere is mounted on a translational stage to move along $z$ direction. Another set of translational stages are used to align it precisely along the $x$--$y$ direction with the hole at the cavity center. The bias magnetic field is applied along $z$ direction by a permanent magnet. The magnet is also mounted on a translations stage that moves along $z$ direction, which tunes the bias magnetic field to the YIG sphere by varying the distances between the magnet and the sphere. During the measurement, a low VNA output power of -5 dBm is used to prevent any magnon nonlinearity effects.

%%%%%%%%%%%%%%%%%%%%%%%%%%%%%%%%%%%%%%%%%%%%%%%%%%%%%%%%%%%%%
%%%%%%%%%%%%%%%%%%%%%%%%%%%%%%%%%%%%%%%%%%%%%%%%%%%%%%%%%%%%%
%%%%%%%%%%%%%%                          %%%%%%%%%%%%%%%%%%%%%
%%%%%%%%%%%%%%         Theory           %%%%%%%%%%%%%%%%%%%%%
%%%%%%%%%%%%%%                          %%%%%%%%%%%%%%%%%%%%%
%%%%%%%%%%%%%%%%%%%%%%%%%%%%%%%%%%%%%%%%%%%%%%%%%%%%%%%%%%%%%
%%%%%%%%%%%%%%%%%%%%%%%%%%%%%%%%%%%%%%%%%%%%%%%%%%%%%%%%%%%%%
\section{Theory derivation}

%%%%%%%%%%%%%%%%%%%%%%%%%%%%%%%%%%%%%%%%%%%%%%%%%%%%%%%%%%%%%
%%%%%%%%%%%%%%%%%%%%%%%%%%%%%%%%%%%%%%%%%%%%%%%%%%%%%%%%%%%%%

\subsection{Equation of Motion}

In our system, the equation of motion of the intra-cavity field a can be generally written as
\begin{equation}
	\dot{\boldsymbol{a}}={A}\boldsymbol{a}+{B}\boldsymbol{s}_\mathrm{in}\label{eq:1.1}
\end{equation}
with input-output relation
\begin{equation}
	\boldsymbol{s}_\mathrm{out}={C}\boldsymbol{s}_\mathrm{in}+{D}\boldsymbol{a}.\label{eq:1.2}
\end{equation}
Here $\boldsymbol{a}=\left(\begin{array}{c}
a\\
b\\
m
\end{array}\right)$, input $\boldsymbol{s}_\mathrm{in}=\left(\begin{array}{c}
s_\mathrm{in1}\\
s_\mathrm{in2}
\end{array}\right)$, and output $\boldsymbol{s}_\mathrm{out}=\left(\begin{array}{c}
s_\mathrm{out1}\\
s_\mathrm{out2}
\end{array}\right)$. $a$, $b$, and $m$ denote the annihilation operators of the two
microwave modes and the magnon mode with respective resonant frequency
$\omega_{a}$, $\omega_{b}$, and $\omega_{m}$.

$A,B,C,D$ are
matrices that determined by the mode interactions and port couplings;
once they are given, one can calculate the scattering matrix $S[\omega]$,
defined by $\boldsymbol{s}_\mathrm{out}[\omega]=S[\omega]\boldsymbol{s}_\mathrm{in}[\omega]$,
by solving Eq.$\,$(\ref{eq:1.1}) and (\ref{eq:1.2}) in the frequency
domain
\begin{equation}
	{S}[\omega]={C}+{D}\left[-i\omega I-A\right]^{-1}B.
\end{equation}
Based on this equation, the system non-reciprocity can be studied by solving $S_{12}$ and
$S_{21}$, which correspond to the transmission spectra measured in
experiment.

To obtain the general forms of $A,B,C,D$, several restrictions have to be taken into consideration. Because of energy conservation, we have $D^{\dagger}D=BB^{\dagger}=\left(\begin{array}{cc}
\kappa_{ae} & 0\\
0 & \kappa_{be}
\end{array}\right)$ and $-C^{\dagger}D=B^{\dagger}$
with $C$ being unitary $C^{\dagger}{C=I}$.
Meanwhile, ${C}$ should be symmetric since the
port response should be reciprocal when not coupled to the cavity
system. Under these restrictions, the general forms of these matrices can be given as

\[
A=\left(\begin{array}{ccc}
-i\omega_{a}-\frac{\kappa_{a}}{2} & 0 & ig_{a}\\
0 & -i\omega_{b}-\frac{\kappa_{b}}{2} & ig_{b}\\
ig_{a} & ig_{b} & -i\omega_{m}-\frac{\kappa_{m}}{2}
\end{array}\right),
\]

\[
B=\left(\begin{array}{cc}
\sqrt{\eta\kappa_{ae}} & \sqrt{(1-\eta)\kappa_{ae}}e^{i\alpha}\\
\sqrt{(1-\eta)\kappa_{be}}e^{i\beta} & \sqrt{\eta\kappa_{be}}e^{i(\alpha+\beta+\pi)}\\
0 & 0
\end{array}\right),
\]

\[
C=\left(\begin{array}{cc}
\sqrt{1-\xi} & i\sqrt{\xi}\\
i\sqrt{\xi} & \sqrt{1-\xi}
\end{array}\right),
\]
\[
D=-CB^{\dagger}.
\]

\noindent where $g_{a}$ and $g_{b}$ are the coupling rates between the magnon mode
and the two orthogonal microwave modes, $\kappa_{a}$, $\kappa_{b}$,
and $\kappa_{m}$ are the total dissipation rates of the three modes,
and $\kappa_{ae},\kappa_{be}$ as the total external coupling rates of
the microwave modes. $0\leq\xi\leq1$ represents the cross talk between the two ports with $\xi=0$ meaning no cross talk while $\xi=1$ meaning maximum cross talk. $0\leq\eta\leq1$ represents the portion of each individual port coupling rate over the total external coupling rate $\kappa_{ae}$ ($\kappa_{be}$) for a single cavity mode. $0\leq(\alpha,\beta)<2\pi$ represent the coupling phase.

In general, non-reciprocity can be achieved by engineering different modes to have different properties when propagating in different directions. This can be done by, for example, exploiting  different couplings between a magnon mode and microwave modes with different chiralities. Furthermore, proper relative excitation phases ($\alpha$ and $\beta$ in ${B}$) need be carefully designed.

%%%%%%%%%%%%%%%%%%%%%%%%%%%%%%%%%%%%%%%%%%%%%%%%%%%%%%%%%%%%%
%%%%%%%%%%%%%%%%%%%%%%%%%%%%%%%%%%%%%%%%%%%%%%%%%%%%%%%%%%%%%

\subsection{Non-degenerate Microwave Cavity Modes}

In the rectangular SIW cavity design ($N_x=10$, $N_y=9$), the two standing microwave modes resonate at different frequencies ($\omega_{a}\neq\omega_{b}$). Because the cavity aspect ratio is close to one, the two modes have similar mode profiles, which also lead to similar mode overlap with the magnon mode. Therefore, here we assume $g_{b}=g_{a}$ and $\kappa_a=\kappa_b$ for simplicity. For standing-wave modes, the excitation phase does not depend on the port position; hence $\alpha\approx0$. For each port, both cavity modes can be simultaneously excited with different phases. Our simulation reveals that at each port, when $\mathrm{TE}_{210}$ mode shows a maximum electric field, $\mathrm{TE}_{120}$ shows
a maximum magnetic field, indicating a $\frac{\pi}{2}$ phase difference between the two modes ($\beta\approx\frac{\pi}{2})$. To simplify the analysis, we also assume each cavity mode couples with both ports equally ($\eta=1/2$). With these assumptions, the calculated transmission ($S_{12}$ and $S_{21}$) for the non-degenerate cavity can be expressed as
\begin{widetext}
\begin{equation}
	S_{ij}=i\frac{2g_a^2\left[i\sqrt{\xi}\frac{\kappa_a}{2}-i(\sqrt{\xi}\pm\sqrt{1-\xi})\frac{\kappa_{ae}}{2}+\sqrt{\xi}\Delta_a\right]
		-\sqrt{\xi}\left[i\frac{\kappa_a}{2}-(i\kappa_{ae}-\Delta_a)\right](i\frac{\kappa_{a}}{2}+\Delta_a)(i\frac{\kappa_{m}}{2}+\Delta_m)}{(i\frac{\kappa_a}{2}+\Delta_a)
		\left[2g_a^2-(i\frac{\kappa_a}{2}+\Delta_a)(i\frac{\kappa_m}{2}+\Delta_m)\right]},
\end{equation}
\end{widetext}
\noindent where $\Delta_a=\omega-\omega_a$ and $\Delta_m=\omega-\omega_m$ are the detuning of the photon and magnon mode, respectively. The $\pm$ sign changes with the signal propagation direction.

If the two modes are well separated from each other ($|\omega_{a}-\omega_{b}|\gg\kappa_{a},\kappa_{b}$),
they can be treated as independent modes. Obviously, when the magnon
mode is on-resonance with one of the microwave modes, the system is reciprocal. However, in our experiments, due to the finite linewidth of the cavity modes, when one mode is excited on resonance, a small portion of the other mode is also excited with a phase difference $\beta\approx\frac{\pi}{2}$.
Effectively, the photon polarization inside the cavity becomes slightly elliptical instead of being purely linear. As a result, weak non-reciprocity is observed when the magnon is on-resonance with one of the microwave modes. In the middle of the two microwave resonances ($\omega=(\omega_a+\omega_b)/2$), both cavity modes are excited with comparable strengths, which when combined give circularly polarized microwave photons. Depending on the excitation port that is used, the chirality can be reversed. Therefore, when magnon is tuned around that frequency, maximum non-reciprocity can manifest.

\subsection{Degenerate Microwave Modes}

\subsubsection{Linear Mode Picture}

For a square cavity ($N_x=N_y$), the two cavity modes are degenerate, giving $\omega_{a}=\omega_{b}$,
$\kappa_{a}=\kappa_{b}$, $\kappa_{ae}=\kappa_{be}$, $g_{a}=g_{b}$. Here we still assume each mode couple with both ports equally ($\eta=\frac{1}{2}$).
In the linearly polarized standing-wave basis ($\alpha=0,\beta=\frac{\pi}{2}$),
matrices $A$ and $B$ can be rewritten as
\[
{A}_\mathrm{lin}=\left(\begin{array}{ccc}
-i\omega_{a}-\frac{\kappa_{a}}{2} & 0 & ig_{a}\\
0 & -i\omega_{a}-\frac{\kappa_{a}}{2} & ig_{a}\\
ig_{a} & ig_{a} & -i\omega_{m}-\frac{\kappa_{m}}{2}
\end{array}\right),
\]
\[
{B}_\mathrm{lin}=\sqrt{\frac{\kappa_{ae}}{2}}\left(\begin{array}{cc}
1 & 1\\
i & -i\\
0 & 0
\end{array}\right).
\]
Accordingly, the transmission can be calculated as
\begin{equation}
	S_{ij}=\pm\frac{ig_a^2\kappa_{ae}}{(\frac{\kappa_a}{2}-i\Delta_a)\left[2g_a^2+(\frac{\kappa_a}{2}-i\Delta_a)(\frac{\kappa_m}{2}-i\Delta_m)\right]},
\end{equation}
\noindent where $i,j=1,2$ and $i\ne j$.

\subsubsection{Circular Mode Picture}

The non-reciprocity can be better understood if circularly polarized 
modes are used as bases. Since the two cavity modes are degenerate, different
eigenmodes can be chosen through a rotation of the eigenbases. In this particular case, a unitary transform ($U$) can be performed
within the microwave subspace
\[
U=\left(\begin{array}{ccc}
\frac{1}{\sqrt{2}} & \frac{1}{\sqrt{2}} & 0\\
\frac{1}{\sqrt{2}} & -\frac{1}{\sqrt{2}} & 0\\
0 & 0 & 1
\end{array}\right).
\]
The resulting new eigenmodes $\tilde{a}=\frac{1}{\sqrt{2}}(a+b),\tilde{b}=\frac{1}{\sqrt{2}}(a-b)$
are two circularly polarized modes. On this circular-mode basis, matrices
${A}$ and ${B}$ become
\[
{A}_\mathrm{cir}=U{A}_\mathrm{lin}U^{\dagger}=\left(\begin{array}{ccc}
-i\omega_{a}-\frac{\kappa_{a}}{2} & 0 & i\sqrt{2}g_{a}\\
0 & -i\omega_{a}-\frac{\kappa_{a}}{2} & 0\\
i\sqrt{2}g_{a} & 0 & -i\omega_{m}-\frac{\kappa_{m}}{2}
\end{array}\right),
\]
\[
{B}_\mathrm{cir}=U{B}_\mathrm{lin}=\frac{\sqrt{\kappa_{ae}}}{2}\left(\begin{array}{cc}
1+i & 1-i\\
1-i & 1+i\\
0 & 0
\end{array}\right).
\]

Clearly in this case only one circular mode ($\tilde{a}$) couples with the
magnon mode, which agrees with the polarization selection rule as described at the beginning of the main text.

\bibliographystyle{apsrev4-1}
%\bibliography{NRSC}

%

\end{document}